\documentclass[aps,english,showpacs,twocolumn]{revtex4-1}
\usepackage{amssymb}
\usepackage{amsmath}
\usepackage{graphicx}
\usepackage{epsfig}

\begin{document}

\title{Scattering properties of anti-parity-time symmetric non-Hermitian
system}
\author{L. Jin}
\email{jinliang@nankai.edu.cn}
\affiliation{School of Physics, Nankai University, Tianjin 300071, China}

\begin{abstract}
{We investigate the scattering properties of an anti-parity-symmetric
non-Hermitian system. The anti-parity-symmetric scattering center possesses
imaginary nearest-neighbor hoppings and real on-site potentials, it has been
experimentally realized through dissipative coupling and frequency detuning
between atomic spin waves. We find that such anti-parity-symmetric system
displays three salient features: Firstly, the reflection and transmission
are both reciprocal. Secondly, the reflection and transmission probabilities
satisfy $R\pm T=1$, which depends on the parity of the scattering center
size. Thirdly, the scattering matrix satisfies $\left( S\sigma _{\mathrm{z}%
}\right) \left( S\sigma _{\mathrm{z}}\right) ^{\ast }=I$ for scattering
center with even-site; for scattering center with odd-site, the dynamics
exhibits Hermitian scattering behavior, possessing unitary scattering matrix
$SS^\dagger=I$.}
\end{abstract}

\maketitle


\section{Introduction}

The concept of parity-time ($\mathcal{PT}$) symmetry has been raised for
more than two decades, researchers are interested in the peculiar effects
caused by $\mathcal{PT}$ symmetry in non-Hermitian systems~\cite%
{Bender98,Dorey01,AM02,Heiss,Jones,Znojil,PRL08,Klaiman,Bendix,JL,Joglekar10,SLonghi}%
. The $\mathcal{PT}$ symmetry breaking was demonstrated in coupled passive
optical waveguides with different losses~\cite{AGuo}. Applied pump beam to
one waveguide, an active $\mathcal{PT}$-symmetric system was realized, the
light power oscillation in exact $\mathcal{PT}$-symmetric phase was observed~%
\cite{CERuter}. In 2014, $\mathcal{PT}$ symmetry was first experimentally
demonstrated in coupled optical microcavities~\cite{BPeng}. The gain is
induced by lasing from the doped $\mathrm{Er}^{3+}$ ions under pumping.
Single mode operation after selectively breaking the $\mathcal{PT}$ symmetry
enhances the mode gain~\cite{LFengPTlasing,HodaeiPTlasing}. The modes are
chiral at exceptional point and lasing directional is controllable~\cite%
{LasingPNAS}. Recently, the enhancement of sensing has been demonstrated
near the exceptional points of $\mathcal{PT}$-symmetric systems.~\cite%
{PTSensingThree,PTSensingTwo}.

Symmetry in physical systems usually leads to symmetric physical properties.
$\mathcal{PT}$ symmetry induces reciprocal scattering~\cite%
{Cannata,Kalish,Ahmed,Mostafazadeh,SChen}. Reflection $\mathcal{PT}$
symmetry protects the reciprocal transmission; axial $\mathcal{PT}$ symmetry
protects the reciprocal reflection~\cite{LXQ,JLPT}. In the presence of
non-Hermiticity, the scattering is not unitary in general situation; leading
to nonreciprocal reflection (transmission) for a reciprocal transmission
(reflection). $\mathcal{PT}$ symmetry and non-Hermiticity are the key points
of the nonreciprocal scattering behavior exhibited in $\mathcal{PT}$%
-symmetric system. Many intriguing phenomena have been observed such as
coherent perfect absorption~\cite{YDChong,CPAScience,CPAHChen,CPAReview},
unidirectional invisibility, reflectionless~\cite{ZLin,LFengNatMater,Alu},
and spectral singularity~\cite{USS}. Until now, the scattering properties of
system with $\mathcal{PT}$ symmetry are explicit; however, anti-$\mathcal{PT}
$ symmetry as a counterpart of $\mathcal{PT}$ symmetry is rarely
investigated~\cite{AAS,AntiPT,LGe13,JHWU14,JHWU15,VVK}. Recently, the
imaginary coupling is experimentally realized through dissipative coupling
between atomic vapors. The system is non-Hermitian and satisfies anti-$%
\mathcal{PT}$ symmetry. The phase-transition threshold and reflectionless
light prorogation have been observed in high resolution~\cite{AntiPT}.

In this paper, inspiring by the experimentally realized anti-$\mathcal{PT}$%
-symmetric system, we study the scattering properties of an anti-$\mathcal{PT%
}$-symmetric non-Hermitian system, which has imaginary couplings and real
on-site potentials. We demonstrate that the reflection and transmission are
both reciprocal. Besides, the difference or summation between the reflection
and transmission probabilities is unity, this relation depends on the parity
of the scattering center. The scattering matrix satisfies $\left( S\sigma _{%
\mathrm{z}}\right) \left( S\sigma _{\mathrm{z}}\right) ^{\ast }=I$ or $%
SS^{\dagger }=I$ for the scattering center with even- or odd-site,
respectively. In the later case, the anti-$\mathcal{PT}$-symmetric
non-Hermitian system exhibits Hermitian scattering behavior.

The remainder of the paper is as follows. In Sec.~\ref{sec_model}, the
system is modelled. In Sec.~\ref{sec_formalism}, the scattering properties
of an anti-$\mathcal{PT}$-symmetric non-Hermitian system is demonstrated. In
Sec.~\ref{sec_illus}, two concrete examples are presented as illustration.
The results are summarized and discussed in Sec.~\ref{sec_summary}.

\section{Model}

\label{sec_model} Recently, anti-$\mathcal{PT}$-symmetric non-Hermitian
system has been realized in atomic vapors~\cite{AntiPT}. Novel coupling
mechanism leads to a dissipative coupling between two atomic spin waves. In
its Hamiltonian, the dissipative coupling is the imaginary coupling and the
detuning between two atomic spin waves is the on-site potential. In this
work, we study the scattering properties of an anti-$\mathcal{PT}$-symmetric
scattering center, which is a tight-binding chain with imaginary couplings
and real on-site potentials. The Hamiltonian of the scattering center reads
\begin{equation}
H_{\mathrm{c}}=\sum_{j=1}^{N}i\kappa _{j}\left( \left\vert j\right\rangle _{%
\mathrm{cc}}\left\langle j+1\right\vert +\left\vert j+1\right\rangle _{%
\mathrm{cc}}\left\langle j\right\vert \right) +V_{j}\left\vert
j\right\rangle _{\mathrm{cc}}\left\langle j\right\vert ,
\end{equation}%
where the couplings satisfy $\kappa _{j}=\kappa _{N+1-j}$ and the on-site
potentials satisfy $V_{j}=-V_{N+1-j}$. $\left\vert j\right\rangle _{\mathrm{c%
}}$ is the basis of the scattering center site-$j$. The parity operator $%
\mathcal{P}$ is defined as the space reflection $\mathcal{P}j\mathcal{P}%
^{-1}=N+1-j$; $\mathcal{T}$ is defined as the time reversal operator $%
\mathcal{T}i\mathcal{T}^{-1}=-i$. Under these definitions, the scattering
center $H_{\mathrm{c}}$ possesses anti-$\mathcal{PT}$ symmetry, which
satisfies $\left( \mathcal{PT}\right) H_{\mathrm{c}}\left( \mathcal{PT}%
\right) ^{-1}=-H_{\mathrm{c}}$. Notably, it is interesting that the anti-$%
\mathcal{PT}$-symmetric Hamiltonian $H_{\mathrm{c}}$ satisfies $\left(
\mathcal{PT}\right) \left( \pm iH_{\mathrm{c}}\right) \left( \mathcal{PT}%
\right) ^{-1}=\pm iH_{\mathrm{c}}$, which indicates that Hamiltonians $\pm
iH_{\mathrm{c}}$ are $\mathcal{PT}$-symmetric.

The input and output leads are connected to the scattering center. The
Hamiltonian of the system is in the form of $H=H_{-}+H_{+}+H_{\mathrm{c}}+H_{%
\mathrm{in}}$, where
\begin{equation}
H_{\pm }=J\sum_{j=\pm 1}^{\pm \infty }(\left\vert j\pm 1\right\rangle _{%
\mathrm{ll}}\left\langle j\right\vert +\mathrm{h.c.}),
\end{equation}%
are the input and output leads with uniform coupling strength $J$. $%
\left\vert j\right\rangle _{\mathrm{l}}$ is the basis of the leads site-$j$.
\begin{equation}
H_{\mathrm{in}}=g\left\vert -1\right\rangle _{\mathrm{lc}}\left\langle
1\right\vert +g\left\vert N\right\rangle _{\mathrm{cl}}\left\langle
+1\right\vert +\mathrm{h.c.,}
\end{equation}%
is the connection Hamiltonian. $\left\vert 1\right\rangle _{\mathrm{c}}$ and
$\left\vert N\right\rangle _{\mathrm{c}}$ are the sites of the scattering
center $H_{\mathrm{c}}$ that connected to the input and output leads $H_{-}$
and $H_{+}$, respectively.

\begin{figure}[tb]
\includegraphics[ bb=45 400 550 640, width=8.6 cm, clip]{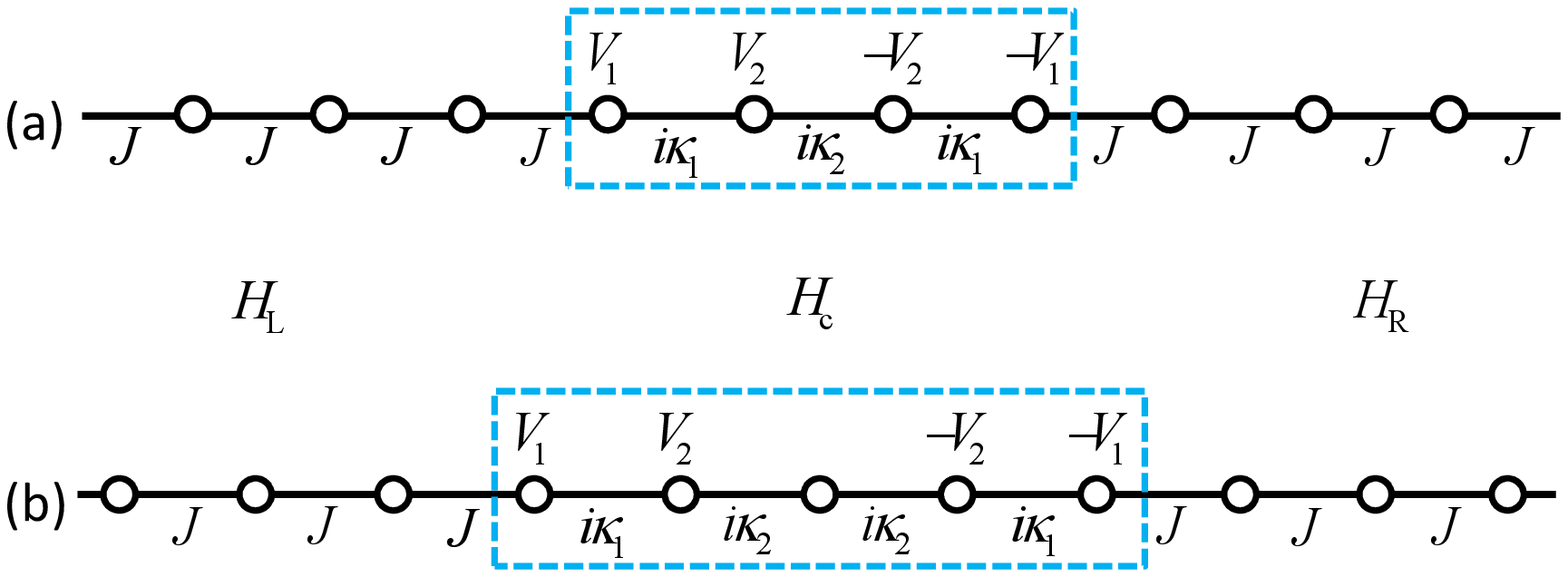}
\caption{(Color online) Schematic illustration of an anti-$\mathcal{PT}$-symmetric scattering system. The site number of the
scattering center is even in (a) and odd in (b).} \label{fig1}
\end{figure}

\section{Scattering formalism}

\label{sec_formalism} In this section, we investigate the scattering
properties of an anti-$\mathcal{PT}$-symmetric non-Hermitian scattering
center, typical scattering behaviors are revealed. In the following, we
discuss the scattering properties of the anti-$\mathcal{PT}$-symmetric
scattering center through investigating the reflection and transmission of
the left and the right inputs. The wave function for the left input is
denoted as $\psi _{\mathrm{L}}^{k}\left( j\right) $ and for the right input
is denoted as $\psi _{\mathrm{R}}^{k}\left( j\right) $ for site $\left\vert
j\right\rangle _{\mathrm{c}}$, where $k$ is the wave vector. The wave
functions are in the form of
\begin{eqnarray}
\psi _{\mathrm{L}}^{k}\left( j\right) &=&\left\{
\begin{array}{c}
e^{ikj}+r_{\mathrm{L}}e^{-ikj},j<0 \\
t_{\mathrm{L}}e^{ikj},j>0%
\end{array}%
\right. ,  \label{Psi_L} \\
\psi _{\mathrm{R}}^{k}\left( j\right) &=&\left\{
\begin{array}{c}
t_{\mathrm{R}}e^{-ikj},j<0 \\
e^{-ikj}+r_{\mathrm{R}}e^{ikj},j>0%
\end{array}%
\right. .  \label{Psi_R}
\end{eqnarray}%
where $r_{\mathrm{L}}$ ($t_{\mathrm{L}}$) and $r_{\mathrm{R}}$ ($t_{\mathrm{R%
}}$) are the reflection (transmission) coefficients for the left and right
inputs, respectively.

\subsection{Identical transmission of transpose invariant}

The scattering center satisfies transpose invariant, i.e., $H_{\mathrm{c}%
}=H_{\mathrm{c}}^{T}$. This leads to identical left and right transmission
coefficients, i.e., $t_{\mathrm{L}}=t_{\mathrm{R}}$. For the left input, the
Schr\"{o}dinger equations for the scattering center are in the form of
\begin{equation}
\left( H_{\mathrm{c}}-EI_{N\times N}\right) \Psi _{\mathrm{c,L}}=\Phi _{%
\mathrm{c,L}},  \label{Delta}
\end{equation}%
where $E=2J\cos k$ is the dispersion relation obtained from the Schr\"{o}%
dinger equations for the leads; $I_{N\times N}$ is the $N\times N$ dimension
identical matrix. $\Psi _{\mathrm{c,L}}$ and $\Phi _{\mathrm{c,L}}$ are $N$
dimension column vectors, their elements are $\Psi _{\mathrm{c,L}}\left(
j\right) =\psi _{\mathrm{c}}^{k}\left( j\right) $ for $j\in \left[ 1,N\right]
$. $\psi _{\mathrm{c}}^{k}\left( j\right) $ represents the wave function of
site-$j$ in the scattering center $H_{\mathrm{c}}$. $\Phi _{\mathrm{c,L}%
}\left( 1\right) =-g\psi _{\mathrm{L}}^{k}\left( -1\right) $, $\Phi _{%
\mathrm{c,L}}\left( N\right) =-g\psi _{\mathrm{L}}^{k}\left( +1\right) $,
and $\Phi _{\mathrm{c,L}}\left( j\right) =0$ for $j\in \left[ 2,N-1\right] $%
. The wave functions at site $\pm 1$ are $\psi _{\mathrm{L}}^{k}\left(
-1\right) =e^{-ik}+r_{\mathrm{L}}e^{ik}$ and $\psi _{\mathrm{L}}^{k}\left(
+1\right) =t_{\mathrm{L}}e^{ik}$. From Eq. (\ref{Delta}), we have%
\begin{eqnarray}
\Psi _{\mathrm{c,L}}\left( 1\right) &=&\Delta _{11}^{-1}\Phi _{\mathrm{c,L}%
}\left( 1\right) +\Delta _{1N}^{-1}\Phi _{\mathrm{c,L}}\left( N\right) , \\
\Psi _{\mathrm{c,L}}\left( N\right) &=&\Delta _{N1}^{-1}\Phi _{\mathrm{c,L}%
}\left( 1\right) +\Delta _{NN}^{-1}\Phi _{\mathrm{c,L}}\left( N\right) ,
\end{eqnarray}%
where $\Delta =H_{\mathrm{c}}-EI_{N\times N}$, and $\Delta _{mn}^{-1}$
represents the element of matrix $\Delta ^{-1}$ on the $m$ row and $n$
column. Then, we have
\begin{eqnarray}
-\frac{\psi _{\mathrm{c}}^{k}\left( 1\right) }{g} &=&\Delta _{11}^{-1}\left(
e^{-ik}+r_{\mathrm{L}}e^{ik}\right) +\Delta _{1N}^{-1}t_{\mathrm{L}}e^{ik},
\\
-\frac{\psi _{\mathrm{c}}^{k}\left( N\right) }{g} &=&\Delta _{N1}^{-1}\left(
e^{-ik}+r_{\mathrm{L}}e^{ik}\right) +\Delta _{NN}^{-1}t_{\mathrm{L}}e^{ik},
\end{eqnarray}

The Schr\"{o}dinger equations for the lead sites $\left\vert -1\right\rangle
_{\mathrm{l}}$ and $\left\vert +1\right\rangle _{\mathrm{l}}$ yield
\begin{eqnarray}
J\psi _{\mathrm{L}}^{k}\left( -2\right) +g\psi _{\mathrm{c}}^{k}\left(
1\right) &=&E\psi _{\mathrm{L}}^{k}\left( -1\right) , \\
J\psi _{\mathrm{L}}^{k}\left( +2\right) +g\psi _{\mathrm{c}}^{k}\left(
N\right) &=&E\psi _{\mathrm{L}}^{k}\left( 1\right) ,
\end{eqnarray}%
the wave functions at sites $\pm 2$ are $\psi _{\mathrm{L}}^{k}\left(
-2\right) =e^{-2ik}+r_{\mathrm{L}}e^{2ik}$ and $\psi _{\mathrm{L}}^{k}\left(
+2\right) =t_{\mathrm{L}}e^{2ik}$. Then, we have
\begin{equation}
\psi _{\mathrm{c}}^{k}\left( 1\right) =\frac{J}{g}\left( 1+r_{\mathrm{L}%
}\right) ,\psi _{\mathrm{c}}^{k}\left( N\right) =\frac{J}{g}t_{\mathrm{L}},
\end{equation}%
the two kinds of expressions for $\psi _{\mathrm{c}}^{k}\left( 1\right) $
and $\psi _{\mathrm{c}}^{k}\left( N\right) $ are equivalent, therefore%
\begin{eqnarray}
-\frac{J\left( 1+r_{\mathrm{L}}\right) }{g^{2}} &=&\Delta _{11}^{-1}\left(
e^{-ik}+r_{\mathrm{L}}e^{ik}\right) +\Delta _{1N}^{-1}t_{\mathrm{L}}e^{ik},
\\
-\frac{J}{g^{2}}t_{\mathrm{L}} &=&\Delta _{N1}^{-1}\left( e^{-ik}+r_{\mathrm{%
L}}e^{ik}\right) +\Delta _{NN}^{-1}t_{\mathrm{L}}e^{ik},
\end{eqnarray}%
and the transmission for the left input is%
\begin{equation}
t_{\mathrm{L}}=\frac{2i\left( J/g^{2}\right) \Delta _{N1}^{-1}\sin k}{\left[
\frac{J}{g^{2}}+\Delta _{NN}^{-1}e^{ik}\right] \left[ \frac{J}{g^{2}}+\Delta
_{11}^{-1}e^{ik}\right] -\Delta _{N1}^{-1}\Delta _{1N}^{-1}e^{2ik}}.
\label{tL}
\end{equation}

For the right input, the Schr\"{o}dinger equations for the scattering center
are in the form of
\begin{equation}
\Delta \Psi _{\mathrm{c,R}}=\Phi _{\mathrm{c,R}},
\end{equation}%
$\Psi _{\mathrm{c,R}}$ and $\Phi _{\mathrm{c,R}}$ are $N$ dimension column
vectors, their elements are $\Psi _{\mathrm{c,R}}\left( j\right) =\psi _{%
\mathrm{c}}^{k}\left( j\right) $ for $j\in \left[ 1,N\right] $; $\Phi _{%
\mathrm{c,R}}\left( 1\right) =-g\psi _{\mathrm{R}}^{k}\left( -1\right) $, $%
\Phi _{\mathrm{c,R}}\left( N\right) =-g\psi _{\mathrm{R}}^{k}\left(
+1\right) $, and $\Phi _{\mathrm{c,R}}\left( j\right) =0$ for $j\in \left[
2,N-1\right] $. The wave functions at sites $\pm 1$ are $\psi _{\mathrm{R}%
}^{k}\left( -1\right) =t_{\mathrm{R}}e^{ik}$ and $\psi _{\mathrm{R}%
}^{k}\left( +1\right) =e^{-ik}+r_{\mathrm{R}}e^{ik}$. From Eq.~(\ref{Delta}%
), we have%
\begin{eqnarray}
\Psi _{\mathrm{c,R}}\left( 1\right) &=&\Delta _{11}^{-1}\Phi _{\mathrm{c,R}%
}\left( 1\right) +\Delta _{1N}^{-1}\Phi _{\mathrm{c,R}}\left( N\right) , \\
\Psi _{\mathrm{c,R}}\left( N\right) &=&\Delta _{N1}^{-1}\Phi _{\mathrm{c,R}%
}\left( 1\right) +\Delta _{NN}^{-1}\Phi _{\mathrm{c,R}}\left( N\right) ,
\end{eqnarray}%
that is
\begin{eqnarray}
-\frac{\psi _{\mathrm{c}}^{k}\left( 1\right) }{g} &=&\Delta _{11}^{-1}t_{%
\mathrm{L}}e^{ik}+\Delta _{1N}^{-1}\left( e^{-ik}+r_{\mathrm{R}%
}e^{ik}\right) , \\
-\frac{\psi _{\mathrm{c}}^{k}\left( N\right) }{g} &=&\Delta _{N1}^{-1}t_{%
\mathrm{L}}e^{ik}+\Delta _{NN}^{-1}\left( e^{-ik}+r_{\mathrm{R}%
}e^{ik}\right) ,
\end{eqnarray}

The Schr\"{o}dinger equations for the lead sites $\left\vert -1\right\rangle
_{\mathrm{l}}$ and $\left\vert +1\right\rangle _{\mathrm{l}}$ yield
\begin{eqnarray}
J\psi _{\mathrm{R}}^{k}\left( -2\right) +g\psi _{\mathrm{c}}^{k}\left(
1\right) &=&E\psi _{\mathrm{R}}^{k}\left( -1\right) , \\
J\psi _{\mathrm{R}}^{k}\left( +2\right) +g\psi _{\mathrm{c}}^{k}\left(
N\right) &=&E\psi _{\mathrm{R}}^{k}\left( +1\right) ,
\end{eqnarray}%
the wave functions at sites $\pm 2$ are $\psi _{\mathrm{R}}^{k}\left(
-2\right) =t_{\mathrm{R}}e^{2ik}$ and $\psi _{\mathrm{R}}^{k}\left(
+2\right) =e^{-2ik}+r_{\mathrm{R}}e^{2ik}$. Then, we have
\begin{equation}
\psi _{\mathrm{c}}^{k}\left( 1\right) =\frac{J}{g}t_{\mathrm{R}},\psi _{%
\mathrm{c}}^{k}\left( N\right) =\frac{J}{g}\left( 1+r_{\mathrm{R}}\right) ,
\end{equation}%
therefore,%
\begin{eqnarray}
-\frac{J}{g^{2}}t_{\mathrm{R}} &=&\Delta _{11}^{-1}t_{\mathrm{L}%
}e^{ik}+\Delta _{1N}^{-1}\left( e^{-ik}+r_{\mathrm{R}}e^{ik}\right) , \\
-\frac{J\left( 1+r_{\mathrm{R}}\right) }{g^{2}} &=&\Delta _{N1}^{-1}t_{%
\mathrm{L}}e^{ik}+\Delta _{NN}^{-1}\left( e^{-ik}+r_{\mathrm{R}%
}e^{ik}\right) ,
\end{eqnarray}%
and the transmission for the right input is%
\begin{equation}
t_{\mathrm{R}}=\frac{2i\left( J/g^{2}\right) \left( \Delta ^{-1}\right)
_{1N}\sin k}{\left[ \frac{J}{g^{2}}+\Delta _{11}^{-1}e^{ik}\right] \left[
\frac{J}{g^{2}}+\Delta _{NN}^{-1}e^{ik}\right] -\Delta _{1N}^{-1}\Delta
_{N1}^{-1}e^{2ik}}.  \label{tR}
\end{equation}

Because $H_{\mathrm{c}}=H_{\mathrm{c}}^{T}$, then we have $\Delta =\Delta
^{T}$. Notice that $\left( \Delta ^{T}\right) ^{-1}=\left( \Delta
^{-1}\right) ^{T}$, then we obtain $\Delta ^{-1}=\left( \Delta ^{T}\right)
^{-1}\Delta ^{T}\Delta ^{-1}=\left( \Delta ^{T}\right) ^{-1}=\left( \Delta
^{-1}\right) ^{T}$. Thus, the matrix elements satisfy $\Delta
_{1N}^{-1}=\Delta _{N1}^{-1}$. Through comparing Eqs. (\ref{tL}) and (\ref%
{tR}), we notice that the left transmission coefficient is identical with
the right transmission coefficient. Therefore, the transpose invariant of $%
H_{\mathrm{c}}$ yields identical transmission coefficients%
\begin{equation}
t_{\mathrm{L}}=t_{\mathrm{R}}.  \label{tLtR}
\end{equation}

\subsection{Reciprocal reflection under $\mathcal{T}$ symmetry}

The scattering center is also invariant under time reversal operation. The
time reversal operator can be expressed as a unitary operator $\mathcal{U}$
multiples the complex conjugation operator $\mathcal{K}$, i.e., $\mathcal{T}=%
\mathcal{U}\mathcal{K}$. The element of the unitary operator $\mathcal{U}$
is $_{\mathrm{c}}\left\langle m\right\vert \mathcal{U}\left\vert
n\right\rangle _{\mathrm{c}}=\left( -1\right) ^{m-1}\delta \left( m-n\right)
$, where $\delta $ is the Dirac delta function.

\begin{figure}[tb]
\includegraphics[ bb=50 370 550 690, width=8.6 cm, clip]{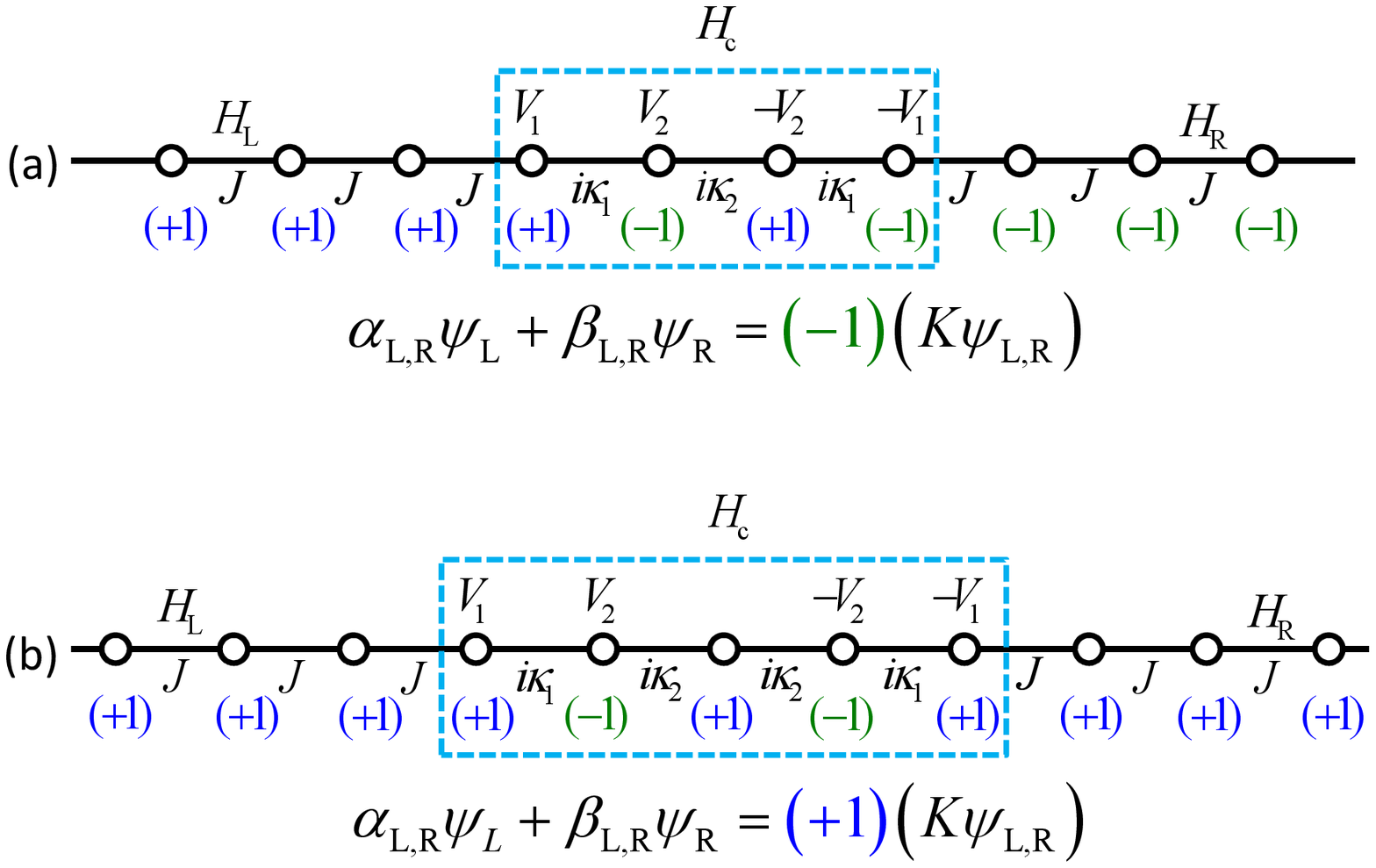}
\caption{(Color online) The wave function relation between the
anti-$\mathcal{PT}$-symmetric scattering system. The site number of the
center is (a) even and (b) odd. The $+1$ in blue and $-1$ in green represent
the unitary transformation, the scattering system changes back to itself after time
reversal operation.} \label{fig2}
\end{figure}

The unitary operator is a diagonal matrix with staggered elements $+1$ and $%
-1$, which is a transformation on the scattering center basis. We
schematically illustrate this basis transformation in Fig.~\ref{fig2} with
the coefficients $+1$ in blue and $-1$ in green. $+1$ indicates that the
basis is unchanged; $-1$ indicates that the basis changes from $\left\vert
j\right\rangle _{\mathrm{c}}$ to $-\left\vert j\right\rangle _{\mathrm{c}}$
after the basis transformation. Figure~\ref{fig2} implies the Hamiltonian of
the scattering center $H_{\mathrm{c}}$ is invariant after the time reversal
operation, i.e., acting the complex conjugation $\mathcal{K}$ and the basis
transformation. To make the whole system Hamiltonian $H$ being invariant
after the time reversal operation, the basis on the left and right leads
need to change accordingly. The basis on the left lead is unchanged, but
changes from $\left\vert j\right\rangle _{\mathrm{l}}$ to $-\left\vert
j\right\rangle _{\mathrm{l}}$ on the right lead for the scattering center
with even-site. To make the whole system Hamiltonian $H$ unchanged after
time reversal operation, the coefficients on the basis of the two leads for
the scattering center with even-site ($N$ is even) are opposite [Fig.~\ref%
{fig2}(a)]; the basis of the two leads for the scattering center with
odd-site ($N$ is odd) is unchanged [Fig.~\ref{fig2}(b)]. This difference
indicates two distinct relations of the scattering wave functions (Fig.~\ref%
{fig2}). The left input and the right input wave functions with identical
wave vector $k$ can compose either the left or the right wave function after
time reversal operation in two alternative ways for the scattering center
with different parities.

We act the complex conjugation operator $\mathcal{K}$ on the wave functions
Eqs. (\ref{Psi_L},\ref{Psi_R}) to get%
\begin{eqnarray}
\mathcal{K}\psi _{\mathrm{L}}^{k}\left( j\right) &=&\left\{
\begin{array}{c}
e^{-ikj}+r_{\mathrm{L}}^{\ast }e^{ikj},j<0 \\
t_{\mathrm{L}}^{\ast }e^{-ikj},j>0%
\end{array}%
\right. , \\
\mathcal{K}\psi _{\mathrm{R}}^{k}\left( j\right) &=&\left\{
\begin{array}{c}
t_{\mathrm{R}}^{\ast }e^{ikj},j<0 \\
e^{ikj}+r_{\mathrm{R}}^{\ast }e^{-ikj},j>0%
\end{array}%
\right. .
\end{eqnarray}%
For the configuration shown in Fig.~\ref{fig2}(a), we can compose $-\mathcal{%
K}\psi _{\mathrm{L}}^{k}\left( j\right) $ in the $j<0$ region through $\psi
_{\mathrm{L}}^{k}\left( j\right) $ and $\psi _{\mathrm{R}}^{k}\left(
j\right) $ of Eqs. (\ref{Psi_L}, \ref{Psi_R}) by eliminating $e^{ikj}$ in $%
j>0$ region. We have
\begin{eqnarray}
&&t_{\mathrm{L}}^{\ast }\left[ \psi _{\mathrm{R}}^{k}\left( j\right) -\frac{%
r_{\mathrm{R}}}{t_{\mathrm{L}}}\psi _{\mathrm{L}}^{k}\left( j\right) \right]
\notag \\
&=&\left\{
\begin{array}{c}
t_{\mathrm{L}}^{\ast }\left( t_{\mathrm{R}}-r_{\mathrm{R}}\frac{r_{\mathrm{L}%
}}{t_{\mathrm{L}}}\right) e^{-ikj}-t_{\mathrm{L}}^{\ast }\frac{r_{\mathrm{R}}%
}{t_{\mathrm{L}}}e^{ikj},j<0 \\
t_{\mathrm{L}}^{\ast }e^{-ikj},j>0%
\end{array}%
\right. ,
\end{eqnarray}%
the coefficients in $j>0$ region for the composed wave function $t_{\mathrm{L%
}}^{\ast }\left[ \psi _{\mathrm{R}}^{k}\left( j\right) -\frac{r_{\mathrm{R}}%
}{t_{\mathrm{L}}}\psi _{\mathrm{L}}^{k}\left( j\right) \right] $ and $%
\mathcal{K}\psi _{\mathrm{L}}^{k}\left( j\right) $ are the same; but they
should be opposite in the $j<0$ region. Therefore, the coefficients in the $%
j<0$ region satisfies
\begin{eqnarray}
t_{\mathrm{L}}^{\ast }\left( t_{\mathrm{R}}-r_{\mathrm{R}}\frac{r_{\mathrm{L}%
}}{t_{\mathrm{L}}}\right) &=&-1, \\
-t_{\mathrm{L}}^{\ast }\frac{r_{\mathrm{R}}}{t_{\mathrm{L}}} &=&-r_{\mathrm{L%
}}^{\ast },
\end{eqnarray}%
then we have the relations%
\begin{equation}
r_{\mathrm{L}}r_{\mathrm{L}}^{\ast }-t_{\mathrm{L}}^{\ast }t_{\mathrm{R}%
}=1;t_{\mathrm{L}}^{\ast }r_{\mathrm{R}}=r_{\mathrm{L}}^{\ast }t_{\mathrm{L}%
},  \label{relationLEven}
\end{equation}%
for the scattering center site number being even.

For the configuration shown in Fig.~\ref{fig2}(b), the composed wave
function $t_{\mathrm{L}}^{\ast }\left[ \psi _{\mathrm{R}}^{k}\left( j\right)
-\frac{r_{\mathrm{R}}}{t_{\mathrm{L}}}\psi _{\mathrm{L}}^{k}\left( j\right) %
\right] $ and $\mathcal{K}\psi _{\mathrm{L}}^{k}\left( j\right) $ are the
same in both the left and the right leads. Then, we obtain
\begin{eqnarray}
t_{\mathrm{L}}^{\ast }\left( t_{\mathrm{R}}-r_{\mathrm{R}}\frac{r_{\mathrm{L}%
}}{t_{\mathrm{L}}}\right) &=&1, \\
-t_{\mathrm{L}}^{\ast }\frac{r_{\mathrm{R}}}{t_{\mathrm{L}}} &=&r_{\mathrm{L}%
}^{\ast },
\end{eqnarray}%
and the relations%
\begin{equation}
r_{\mathrm{L}}r_{\mathrm{L}}^{\ast }+t_{\mathrm{L}}^{\ast }t_{\mathrm{R}%
}=1;t_{\mathrm{L}}^{\ast }r_{\mathrm{R}}=-r_{\mathrm{L}}^{\ast }t_{\mathrm{L}%
},  \label{relationLOdd}
\end{equation}%
for the scattering center site number being odd.

For the configuration shown in Fig.~\ref{fig2}(a), we compose $-\mathcal{K}%
\psi _{\mathrm{R}}^{k}\left( j\right) $ via $\psi _{\mathrm{L}}^{k}\left(
j\right) $ and $\psi _{\mathrm{R}}^{k}\left( j\right) $ of Eqs. (\ref{Psi_L}%
, \ref{Psi_R}) by eliminating $e^{-ikj}$ in $j<0$ region. We have%
\begin{eqnarray}
&&t_{\mathrm{R}}^{\ast }\left[ \psi _{\mathrm{L}}^{k}\left( j\right) -\frac{%
r_{\mathrm{L}}}{t_{\mathrm{R}}}\psi _{\mathrm{R}}^{k}\left( j\right) \right]
\notag \\
&=&\left\{
\begin{array}{c}
t_{\mathrm{R}}^{\ast }e^{ikj},j<0 \\
t_{\mathrm{R}}^{\ast }\left( t_{\mathrm{L}}-r_{\mathrm{L}}\frac{r_{\mathrm{R}%
}}{t_{\mathrm{R}}}\right) e^{ikj}-t_{\mathrm{R}}^{\ast }\frac{r_{\mathrm{L}}%
}{t_{\mathrm{R}}}e^{-ikj},j>0%
\end{array}%
\right. ,
\end{eqnarray}%
the coefficients in $j<0$ region for the composed wave function $t_{\mathrm{R%
}}^{\ast }\left[ \psi _{\mathrm{L}}^{k}\left( j\right) -\frac{r_{\mathrm{L}}%
}{t_{\mathrm{R}}}\psi _{\mathrm{R}}^{k}\left( j\right) \right] $ and $%
\mathcal{K}\psi _{\mathrm{R}}^{k}\left( j\right) $ are identical; but the
coefficients in the $j>0$ region should be opposite. Therefore, we have the
relations%
\begin{eqnarray}
t_{\mathrm{R}}^{\ast }\left( t_{\mathrm{L}}-r_{\mathrm{L}}\frac{r_{\mathrm{R}%
}}{t_{\mathrm{R}}}\right) &=&-1, \\
-t_{\mathrm{R}}^{\ast }\frac{r_{\mathrm{L}}}{t_{\mathrm{R}}} &=&-r_{\mathrm{R%
}}^{\ast }.
\end{eqnarray}%
Simplifying the obtained relations, we have%
\begin{equation}
r_{\mathrm{R}}r_{\mathrm{R}}^{\ast }-t_{\mathrm{R}}^{\ast }t_{\mathrm{L}%
}=1;t_{\mathrm{R}}^{\ast }r_{\mathrm{L}}=r_{\mathrm{R}}^{\ast }t_{\mathrm{R}%
},  \label{relationREven}
\end{equation}%
for the scattering center site number being even.

For the configuration shown in Fig.~\ref{fig2}(b), the composed wave
function $t_{\mathrm{L}}^{\ast }\left[ \psi _{\mathrm{R}}^{k}\left( j\right)
-\frac{r_{\mathrm{R}}}{t_{\mathrm{L}}}\psi _{\mathrm{L}}^{k}\left( j\right) %
\right] $ and $\mathcal{K}\psi _{\mathrm{L}}^{k}\left( j\right) $ are the
same in both the left and the right leads. Thus, we obtain
\begin{eqnarray}
t_{\mathrm{R}}^{\ast }\left( t_{\mathrm{L}}-r_{\mathrm{L}}\frac{r_{\mathrm{R}%
}}{t_{\mathrm{R}}}\right) &=&1, \\
-t_{\mathrm{R}}^{\ast }\frac{r_{\mathrm{L}}}{t_{\mathrm{R}}} &=&r_{\mathrm{R}%
}^{\ast },
\end{eqnarray}%
after simplification, we obtain the relations%
\begin{equation}
r_{\mathrm{R}}r_{\mathrm{R}}^{\ast }+t_{\mathrm{R}}^{\ast }t_{\mathrm{L}%
}=1;t_{\mathrm{R}}^{\ast }r_{\mathrm{L}}=-t_{\mathrm{R}}r_{\mathrm{R}}^{\ast
},  \label{relationROdd}
\end{equation}%
for the scattering center site number being odd.

A direct conclusion from the relations of scattering coefficients Eqs.~(\ref%
{relationLEven},~\ref{relationREven}) and Eqs.~(\ref{relationLOdd},~\ref%
{relationROdd}) is the reciprocal reflection, i.e., $\left\vert r_{\mathrm{L}%
}\right\vert =\left\vert r_{\mathrm{R}}\right\vert $ in both configurations
of Fig.~\ref{fig1}.

\subsection{Scattering probability and scattering matrix}

For the scattering center with even-site, their reflection and transmission
satisfy Eqs. (\ref{tLtR}), (\ref{relationLEven}), and (\ref{relationREven}),
from which we first obtain%
\begin{equation}
\left\vert r_{\mathrm{L}}\right\vert ^{2}-\left\vert t_{\mathrm{L}%
}\right\vert ^{2}=\left\vert r_{\mathrm{R}}\right\vert ^{2}-\left\vert t_{%
\mathrm{R}}\right\vert ^{2}=1.
\end{equation}%
And then, we obtain that the scattering matrix satisfies
\begin{equation}
\left( S\sigma _{\mathrm{z}}\right) \left( S\sigma _{\mathrm{z}}\right)
^{\ast }=\left(
\begin{array}{cc}
r_{\mathrm{L}}r_{\mathrm{L}}^{\ast }-t_{\mathrm{R}}t_{\mathrm{R}}^{\ast } &
t_{\mathrm{R}}r_{\mathrm{R}}^{\ast }-r_{\mathrm{L}}t_{\mathrm{L}}^{\ast } \\
t_{\mathrm{L}}r_{\mathrm{L}}^{\ast }-r_{\mathrm{R}}t_{\mathrm{R}}^{\ast } &
r_{\mathrm{R}}r_{\mathrm{R}}^{\ast }-t_{\mathrm{L}}t_{\mathrm{L}}^{\ast }%
\end{array}%
\right) =I_{2\times 2}\text{,}
\end{equation}%
in the configuration shown in Fig.~\ref{fig2}(a), where $S$ is the
scattering matrix and $\sigma _{\mathrm{z}}$ is the Pauli matrix defined as
\begin{equation}
S=\left(
\begin{array}{cc}
r_{\mathrm{L}} & t_{\mathrm{R}} \\
t_{\mathrm{L}} & r_{\mathrm{R}}%
\end{array}%
\right) ,\sigma _{\mathrm{z}}=\left(
\begin{array}{cc}
1 & 0 \\
0 & -1%
\end{array}%
\right) .
\end{equation}

For the scattering center with odd-site, their reflection and transmission
satisfy Eqs. (\ref{tLtR}), (\ref{relationLOdd}), and (\ref{relationROdd}),
from which we first obtain%
\begin{equation}
\left\vert r_{\mathrm{L}}\right\vert ^{2}+\left\vert t_{\mathrm{L}%
}\right\vert ^{2}=\left\vert r_{\mathrm{R}}\right\vert ^{2}+\left\vert t_{%
\mathrm{R}}\right\vert ^{2}=1.
\end{equation}%
And then, we obtain that the scattering matrix is unitary in the
configuration shown in Fig.~\ref{fig2}(b),
\begin{equation}
SS^{\dagger }=\left(
\begin{array}{cc}
r_{\mathrm{L}}r_{\mathrm{L}}^{\ast }+t_{\mathrm{R}}t_{\mathrm{R}}^{\ast } &
r_{\mathrm{L}}t_{\mathrm{L}}^{\ast }+t_{\mathrm{R}}r_{\mathrm{R}}^{\ast } \\
t_{\mathrm{L}}r_{\mathrm{L}}^{\ast }+r_{\mathrm{R}}t_{\mathrm{R}}^{\ast } &
t_{\mathrm{L}}t_{\mathrm{L}}^{\ast }+r_{\mathrm{R}}r_{\mathrm{R}}^{\ast }%
\end{array}%
\right) =I_{2\times 2}\text{.}
\end{equation}%
The scattering dynamics exhibited in the odd-site anti-$\mathcal{PT}$%
-symmetric scattering center is similar as the dynamics in a Hermitian
scattering center. Therefore, unitary scattering not only occurs in $%
\mathcal{PT}$-symmetric non-Hermitian system~\cite{SChen}, but also appears
in anti-$\mathcal{PT}$-symmetric non-Hermitian system.

\section{Illustrative examples}

\label{sec_illus}

We consider concrete models to demonstrate our results. The two leads are $%
H_{\pm }=-\sum_{j=\pm 1}^{\pm \infty }\left\vert j\pm 1\right\rangle _{%
\mathrm{ll}}\left\langle j\right\vert +\mathrm{H.c.}$; the connection
Hamiltonian is $H_{\mathrm{in}}=-\left\vert -1\right\rangle _{\mathrm{lc}%
}\left\langle 1\right\vert -\left\vert 2\right\rangle _{\mathrm{cl}%
}\left\langle +1\right\vert +\mathrm{H.c.}$; and the Hamiltonian of the
two-site scattering center is%
\begin{equation}
H_{\mathrm{c}}^{\mathrm{2d}}=\left(
\begin{array}{cc}
V & -i \\
-i & -V%
\end{array}%
\right) .
\end{equation}%
The scattering center satisfies $\left( \mathcal{PT}\right) H_{\mathrm{c}}^{%
\mathrm{2d}}\left( \mathcal{PT}\right) ^{-1}=-H_{\mathrm{c}}^{\mathrm{2d}}$.
The Schr\"{o}dinger equations for the scattering center are%
\begin{eqnarray}
-\psi _{\mathrm{L(R)}}^{k}\left( -1\right) -i\psi _{\mathrm{c}}^{k}\left(
2\right) &=&\left( E-V\right) \psi _{\mathrm{c}}^{k}\left( 1\right) , \\
-\psi _{\mathrm{L(R)}}^{k}\left( +1\right) -i\psi _{\mathrm{c}}^{k}\left(
1\right) &=&\left( E+V\right) \psi _{\mathrm{c}}^{k}\left( 2\right) ,
\end{eqnarray}%
the dispersion is $E=-2\cos k$. For the left input, we set the wave
functions as $\psi _{\mathrm{L}}^{k}\left( -1\right) =e^{-ik}+r_{\mathrm{L}%
}e^{ik}$, $\psi _{\mathrm{c}}^{k}\left( 1\right) =1+r_{\mathrm{L}}$, $\psi _{%
\mathrm{c}}^{k}\left( 2\right) =t_{\mathrm{L}}$, and $\psi _{\mathrm{L}%
}^{k}\left( +1\right) =t_{\mathrm{L}}e^{ik}$. For the right input, we set
the wave functions as $\psi _{\mathrm{R}}^{k}\left( -1\right) =t_{\mathrm{R}%
}e^{ik}$, $\psi _{\mathrm{c}}^{k}\left( 1\right) =t_{\mathrm{R}}$, $\psi _{%
\mathrm{c}}^{k}\left( 2\right) =1+r_{\mathrm{R}}$, and $\psi _{\mathrm{R}%
}^{k}\left( +1\right) =e^{-ik}+r_{\mathrm{R}}e^{ik}$. Substituting the wave
functions into the Schr\"{o}dinger equations, we obtain the reflection and
transmission, which read
\begin{eqnarray}
t_{\mathrm{L}} &=&t_{\mathrm{R}}=\frac{2\sin k}{2\cos ke^{-ik}-V^{2}}, \\
r_{\mathrm{L}} &=&\frac{V^{2}-2+2iV\sin k}{2\cos ke^{-ik}-V^{2}}, \\
r_{\mathrm{R}} &=&\frac{V^{2}-2-2iV\sin k}{2\cos ke^{-ik}-V^{2}}.
\end{eqnarray}%
Notably, $t_{\mathrm{L}}=t_{\mathrm{R}}$, $\left\vert r_{\mathrm{L}%
}\right\vert =\left\vert r_{\mathrm{R}}\right\vert $, and $\left\vert r_{%
\mathrm{L}(\mathrm{R})}\right\vert ^{2}-\left\vert t_{\mathrm{L}(\mathrm{R}%
)}\right\vert ^{2}=1$. The scattering matrix satisfies $(S\sigma _{\mathrm{z}%
})(S\sigma _{\mathrm{z}})^{\ast }=I$.

For a three-site anti-$\mathcal{PT}$-symmetric scattering center
\begin{equation}
H_{\mathrm{c}}^{\mathrm{3d}}=\left(
\begin{array}{ccc}
V & -i & 0 \\
-i & 0 & -i \\
0 & -i & -V%
\end{array}%
\right) ,
\end{equation}
we notice that $\left( \mathcal{PT}\right) H_{\mathrm{c}}^{\mathrm{3d}%
}\left( \mathcal{PT}\right) ^{-1}=-H_{\mathrm{c}}^{\mathrm{3d}}$. The
connection Hamiltonian is $H_{\mathrm{in}}=-\left\vert -1\right\rangle _{%
\mathrm{lc}}\left\langle 1\right\vert -\left\vert 3\right\rangle _{\mathrm{cl%
}}\left\langle +1\right\vert +\mathrm{H.c.}$, and the Schr\"{o}dinger
equations are%
\begin{eqnarray}
-\psi _{\mathrm{L(R)}}^{k}\left( -1\right) -i\psi _{\mathrm{c}}^{k}\left(
2\right) &=&\left( E-V\right) \psi _{\mathrm{c}}^{k}\left( 1\right) , \\
-i\psi _{\mathrm{c}}^{k}\left( 1\right) -i\psi _{\mathrm{c}}^{k}\left(
3\right) &=&E\psi _{\mathrm{c}}^{k}\left( 2\right) , \\
-\psi _{\mathrm{L(R)}}^{k}\left( +1\right) -i\psi _{\mathrm{c}}^{k}\left(
2\right) &=&\left( E+V\right) \psi _{\mathrm{c}}^{k}\left( 3\right) .
\end{eqnarray}%
For\ the left input, the wave functions are set as $\psi _{\mathrm{L}%
}^{k}\left( -1\right) =e^{-ik}\left( e^{-ik}+r_{\mathrm{L}}e^{ik}\right) $, $%
\psi _{\mathrm{c}}^{k}\left( 1\right) =e^{-ik}\left( 1+r_{\mathrm{L}}\right)
$, $\psi _{\mathrm{c}}^{k}\left( 3\right) =e^{-ik}t_{\mathrm{L}}$, and $\psi
_{\mathrm{L}}^{k}\left( +1\right) =t_{\mathrm{L}}$. For the right input, the
wave functions are set as $\psi _{\mathrm{R}}^{k}\left( -1\right) =t_{%
\mathrm{R}}$, $\psi _{\mathrm{c}}^{k}\left( 1\right) =e^{-ik}t_{\mathrm{R}}$%
, $\psi _{\mathrm{c}}^{k}\left( 3\right) =e^{-ik}\left( 1+r_{\mathrm{R}%
}\right) $, and $\psi _{\mathrm{R}}^{k}\left( +1\right) =e^{-ik}\left(
e^{-ik}+r_{\mathrm{R}}e^{ik}\right) $. After simplification, we obtain the
reflection and transmission
\begin{eqnarray}
t_{\mathrm{L}} &=&t_{\mathrm{R}}=\frac{i\sin k}{\left( e^{-2ik}-V^{2}\right)
\cos k+e^{-ik}}, \\
r_{\mathrm{L}} &=&-\frac{\left( e^{ik}+V\right) \left( e^{-ik}-V\right) +1}{%
\left( e^{-2ik}-V^{2}\right) \cos k+e^{-ik}}\cos k, \\
r_{\mathrm{R}} &=&-\frac{\left( e^{ik}-V\right) \left( e^{-ik}+V\right) +1}{%
\left( e^{-2ik}-V^{2}\right) \cos k+e^{-ik}}\cos k.
\end{eqnarray}%
Thus, $t_{\mathrm{L}}=t_{\mathrm{R}}$, $\left\vert r_{\mathrm{L}}\right\vert
=\left\vert r_{\mathrm{R}}\right\vert $, and $\left\vert r_{\mathrm{L}(%
\mathrm{R})}\right\vert ^{2}+\left\vert t_{\mathrm{L}(\mathrm{R}%
)}\right\vert ^{2}=1$. The scattering matrix is unitary, i.e., $SS^{\dagger
}=I$, the scattering dynamics is Hermitian-like.

\begin{figure}[tb]
\includegraphics[ bb=0 0 510 250, width=8.6 cm, clip]{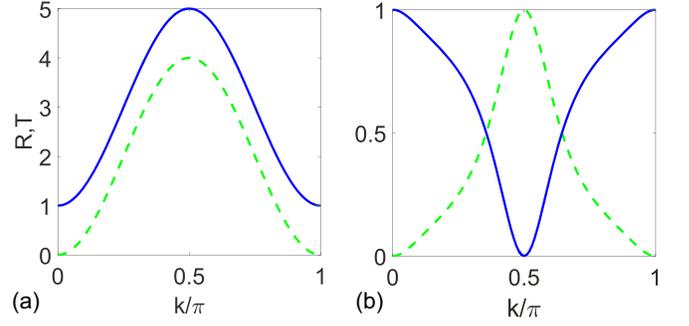}
\caption{(Color online) The reflection (solid blue) and transmission (dashed green)  probabilities as a function of the input wave vector $k$.
Plots are for (a) $H_{\mathrm{c}}^{\mathrm{D}}$ and (b) $H_{\mathrm{c}}^{\mathrm{T}}$ at $V=1$.}
\label{fig3}
\end{figure}

In Fig.~\ref{fig3}, we plot the reciprocal reflection ($R=R_{\mathrm{L}}=R_{%
\mathrm{R}}$) and reciprocal transmission ($T=T_{\mathrm{L}}=T_{\mathrm{R}}$%
) probabilities. In Fig.~\ref{fig3}(a), $R$ and $T$ are both maximal at $%
k=\pi /2$, where the input wave has the largest group velocity. The
reflection and transmission probabilities monotonously increase as $V$
decreases. Both $R$ and $T$ diverge at $k=\pi /2$ when $V=0$, it corresponds
to a spectral singularity and induces symmetric lasing toward both leads~%
\cite{AliSS}. As $V$ increases, the variations on $R$ and $T$ tend to be
flat. In Fig.~\ref{fig3}(b), the reflection is zero and the transmission is
unity at $k=\pi /2$, which corresponds to a resonant transmission that
independent of on-site potentials $V$. As $V$ increases, the variations on $%
R $ and $T$ around $k=\pi /2$ become sharp. Notably, the spectral
singularity can not exist in the discussed anti-$\mathcal{PT}$-symmetric
scattering center with site number being odd, where the scattering exhibits
Hermitian behavior.

\section{Summary and Discussion}

\label{sec_summary}

We have investigated the scattering behavior of an anti-$\mathcal{PT}$%
-symmetric non-Hermitian scattering center $H_{\mathrm{c}}$ with imaginary
nearest-neighbor couplings and real on-site potentials, this type of
scattering centers has the feature that $\left( \pm iH_{\mathrm{c}}\right) $
satisfies the $\mathcal{PT}$ symmetry. We find that the reflection ($R$) and
transmission ($T$) are both reciprocal; and the probabilities satisfy $R\pm
T=1$, which depends on the scattering center size. The scattering matrix of
an even-site scattering center satisfies $\left( S\sigma _{\mathrm{z}%
}\right) \left( S\sigma _{\mathrm{z}}\right) ^{\ast }=I$; an odd-site
scattering center exhibits Hermitian scattering dynamics, its scattering
matrix is unitary $SS^{\dagger }=I$ and none spectral singularity exists. We
would like to state that all the conclusions still valid for the scattering
center with long range imaginary couplings if all the couplings are between
sites with different parity, i.e., only couplings between sites $\left\vert
\mathrm{Odd}\right\rangle _{\mathrm{c}}$ and $\left\vert \mathrm{Even}%
\right\rangle _{\mathrm{c}}$ are nonzero; otherwise, only $t_{\mathrm{L}}=t_{%
\mathrm{R}}$ is valid because of the transpose invariant of the scattering
center. Our results are useful in predicting the propagation features of
anti-$\mathcal{PT}$-symmetric systems and their applications in optics.

\acknowledgments We acknowledge support from NSFC (Grant No. 11605094) and
the Tianjin Natural Science Foundation (Grant No. 16JCYBJC40800).


\begin{thebibliography}{99}
\bibitem{Bender98} C. M. Bender, and S. Boettcher, Phys. Rev. Lett. \textbf{%
80}, 5243 (1998).

\bibitem{Dorey01} P. Dorey, C. Dunning, and R. Tateo, J. Phys. A: Math. Gen.
\textbf{34}, L391 (2001).

\bibitem{AM02} A. Mostafazadeh, J. Math. Phys. \textbf{43}, 3944 (2002).

\bibitem{Heiss} W. D. Heiss, J. Phys. A: Math. Gen. \textbf{37}, 2455 (2004).

\bibitem{Jones} H. F. Jones, J. Phys. A: Math. Gen. \textbf{38}, 1741 (2005).

\bibitem{Znojil} M. Znojil, J. Phys. A: Math. Theor. \textbf{41}, 292002
(2008).

\bibitem{PRL08} K. G. Makris, R. El-Ganainy, D. N. Christodoulides, and Z.
H. Musslimani, Phys. Rev. Lett. \textbf{100}, 103904 (2008).

\bibitem{Klaiman} S. Klaiman, U. G\"{u}nther, and N. Moiseyev, Phys. Rev.
Lett. \textbf{101}, 080402 (2008).

\bibitem{Bendix} O. Bendix, R. Fleischmann, T. Kottos and B. Shapiro, Phys.
Rev. Lett. \textbf{103}, 030402 (2009).

\bibitem{JL} L. Jin and Z. Song, Phys. Rev. A \textbf{80}, 052107 (2009).

\bibitem{SLonghi} S. Longhi, Phys. Rev. A \textbf{82}, 031801(R) (2010).

\bibitem{Joglekar10} Y. N. Joglekar, D. Scott, M. Babbey, and A. Saxena,
Phys. Rev. A \textbf{82}, 030103(R) (2010).

\bibitem{AGuo} A. Guo, G. J. Salamo, D. Duchesne, R.Morandotti, M.
Volatier-Ravat, V. Aimez, G. A. Siviloglou, and D. N. Christodoulides, Phys.
Rev. Lett. \textbf{103}, 093902 (2009).

\bibitem{CERuter} C. E. R\"{u}ter, K. G. Makris, R. El-Ganainy, D. N.
Christodoulides, M. Segev, and D. Kip, Nat. Phys. \textbf{6}, 192 (2010).

\bibitem{BPeng} B. Peng, S. K. \"{O}zdemir, F. Lei, F. Monifi, M. Gianfreda,
G. L. Long, S. Fan, F. Nori, C. M. Bender, and L. Yang, Nat. Phys. \textbf{10%
}, 394 (2014).

\bibitem{LFengPTlasing} L. Feng, Z. J. Wong, R.-M. Ma, Y. Wang, and X.
Zhang, Science \textbf{346}, 972 (2014).

\bibitem{HodaeiPTlasing} H. Hodaei, M.-A. Miri,M. Heinrich, D. N.
Christodoulides, and M. Khajavikhan, Science \textbf{346}, 975 (2014).

\bibitem{LasingPNAS} B. Peng, S. K. \"{O}zdemira, M. Liertzer, W. Chen, J.
Kramer, H. Y\i lmaz, J. Wiersig, S. Rotter, and L. Yang, Proc. Nat. Acad.
Sci. USA \textbf{113}, 6845 (2016).

\bibitem{PTSensingThree} H. Hodaei, A. U. Hassan, S. Wittek, H.
Garcia-Gracia, R. El-Ganainy, D. N. Christodoulides, and M. Khajavikhan,
Nature \textbf{548}, 187 (2017).

\bibitem{PTSensingTwo} W. Chen, S. K. \"{O}zdemir, G. Zhao, J. Wiersig, and
L. Yang, Nature \textbf{548}, 192 (2017).

\bibitem{Cannata} F. Cannata, J.-P. Dedonder, and A. Ventura, Ann. Phys.
(NY) \textbf{322}, 397 (2007).

\bibitem{Kalish} S. Kalish, Z. Lin, and T. Kottos, Phys. Rev. A \textbf{85},
055802 (2012).

\bibitem{Ahmed} Z. Ahmed, Phys. Lett. A \textbf{377}, 957 (2013).

\bibitem{Mostafazadeh} A. Mostafazadeh, J. Phys. A: Math. Theor. \textbf{47}%
, 505303 (2014).

\bibitem{SChen} B. Zhu, R. L\"{u}, and S. Chen, Phys. Rev. A \textbf{91},
042131 (2015).

\bibitem{LXQ} X. Q. Li, X. Z. Zhang, G. Zhang, and Z. Song, Phys. Rev. A
\textbf{91}, 032101 (2015).

\bibitem{JLPT} L. Jin, X. Z. Zhang, G. Zhang, and Z. Song, Sci. Rep. \textbf{%
6}, 20976 (2016).

\bibitem{YDChong} Y. D. Chong, Li Ge, Hui Cao and A. D. Stone, Phys. Rev.
Lett. \textbf{105}, 053901 (2010).

\bibitem{CPAScience} W. Wan, Y. Chong, L. Ge, H. Noh, A. D. Stone, H. Cao,
Science \textbf{331}, 889 (2011).

\bibitem{CPAHChen} Y. Sun, W. Tan, H.-Q. Li, J. Li, H. Chen, Phys. Rev.
Lett. \textbf{112}, 143903 (2014).

\bibitem{CPAReview} D. G. Baranov, A. Krasnok, T. Shegai, A. Al\`{u}, and Y.
Chong, Nat. Rev. Mater. \textbf{2}, 17064 (2017).

\bibitem{ZLin} Z. Lin, H. Ramezani, T. Eichelkraut, T. Kottos, H. Cao, and
D. N. Christodoulides, Phys. Rev. Lett. \textbf{106}, 213901 (2011).

\bibitem{LFengNatMater} L. Feng, Y. L. Xu, W. S. Fegadolli, M. H. Lu, J. E.
B. Oliveira, V. R. Almeida, Y. F. Chen, and A. Scherer, Nat. Mater. \textbf{%
12}, 108 (2013).

\bibitem{Alu} R. Fleury, D. Sounas, and A. Al\`{u}, Nat. Commun. \textbf{6},
5905 (2015).

\bibitem{USS} H. Ramezani , H. K. Li, Y. Wang, and X. Zhang, Phys. Rev.
Lett. \textbf{113}, 263905 (2014).

\bibitem{AAS} D. A. Antonosyan, A. S. Solntsev, and A. A. Sukhorukov, Opt.
Lett. \textbf{40}, 4575 (2015).

\bibitem{AntiPT} P. Peng, W. Cao, C. Shen, W. Qu, J. Wen, L. Jiang, and Y.
Xiao, Nat. Phys. \textbf{12}, 1139 (2016).

\bibitem{LGe13} L. Ge and H. E. T\"{u}reci, Phys. Rev. A \textbf{88}, 053810
(2013).

\bibitem{JHWU14} J.-H. Wu, M. Artoni, and G. C. La Rocca, Phys. Rev. Lett.
\textbf{113}, 123004 (2014).

\bibitem{JHWU15} J.-H. Wu, M. Artoni, and G. C. La Rocca, Phys. Rev. A
\textbf{91}, 033811 (2015).

\bibitem{VVK} V. V. Konotop and D. A. Zezyulin, Phys. Rev. Lett. \textbf{120}%
, 123902 (2018).

\bibitem{AliSS} A. Mostafazadeh, Phys. Rev. Lett. \textbf{102}, 220402
(2009).
\end{thebibliography}
\end{document}